\documentclass{article}

\usepackage{spconfa4}
\usepackage{amsmath,graphicx}
\usepackage{multirow}
\usepackage{booktabs}
\usepackage{xspace}
\usepackage{cite}
\usepackage{enumitem}
\usepackage{pgfplots}
\usepackage{flushend}
\usepackage{url}
\pgfplotsset{compat=newest}
\pgfplotsset{plot coordinates/math parser=false}
\newlength\figureheight
\newlength\figurewidth

\title{Evaluating Speech Enhancement Systems Through Listening Effort}
\name{Femke B. Gelderblom$^1$, Tron V. Tronstad$^1$, and Iv\'an L\'opez-Espejo$^2$\thanks{This work has been supported by the Spanish Ministry of Science and Innovation under the ``Ram\'on y Cajal'' programme (RYC2022-036755-I).}}
\address{$^1$Department of Sustainable Communication Technologies, SINTEF Digital, Norway\\
	$^2$Department of Signal Theory, Telematics and Communications, University of Granada, Spain\\
}
\begin{document}
%\ninept
%
\maketitle
\begin{abstract}
Understanding degraded speech is demanding, requiring increased listening effort (LE). Evaluating processed and unprocessed speech with respect to LE can objectively indicate if speech enhancement systems benefit listeners. However, existing methods for measuring LE are complex and not widely applicable. In this study, we propose a simple method to evaluate speech intelligibility and LE simultaneously without additional strain on subjects or operators. We assess this method using results from two independent studies in Norway and Denmark, testing 76 (50+26) subjects across 9 (6+3) processing conditions. Despite differences in evaluation setups, subject recruitment, and processing systems, trends are strikingly similar, demonstrating the proposed method's robustness and ease of implementation into existing practices.
\end{abstract}
\section{Introduction}
\label{sec:intro}

Speech enhancement (SE) systems aim to improve degraded speech signals and have many important applications within telecommunications, hearing aids, and media production. It therefore comes as no surprise that this is an active field of research, where new systems are proposed frequently. 

Inherent to the development of SE systems is the need for methods to evaluate said systems. The most used evaluation methods are objective metrics that estimate the performance of SE systems based on a measure of the difference between the degraded/processed speech signal and a clean speech reference signal. Popular examples of such metrics are STOI~\cite{Taal2011} for intelligibility and PESQ~\cite{ITU2001} for quality.%, but there are many more to choose from.
%Alternatively, there are also non-intrusive measures available like SRMR~\cite{Falk2010} that do not require a clean reference signal and instead are able to estimate quality or intelligibility of a speech signal based solely on the latter signal itself.
%(and its extended version eSTOI~\cite{Jensen2016})
%I have removed the citation to Taal2010 to save some space.

The advantage of all objective metrics is obvious: they facilitate quick and cost-effective evaluation of SE systems at minimal effort to the developers. However, objective metrics are not without their shortcomings. An objective measure should never be \textit{assumed} to work under conditions different to those it has been validated for. Indeed, several studies report a lack of predictive power of objective intelligibility measures (despite their widespread adaptation), especially for the deep learning-based systems that are currently the main focus of the SE system field of research~\cite{Lopez23, Gelderblom2017, Gelderblom2019, Gelderblom2024, Zhao2018, Healy2015}. This is problematic, because it has long been known that while many SE algorithms can improve quality, they generally do so at the cost of intelligibility~\cite{Lim1979}. 

%Naturally, a SE system should not make the speech signal worse. However, one argument against testing with respect to intelligibility is that, in many of their applications, SE systems act on signals that are degraded, but not to such a degree that intelligibility is affected. In online meetings, for example, the audio might not be considered to be ideal, but it is generally possible to understand what is being said.

With its main focus on the \textit{quality} of speech signals, and marginal focus on \textit{intelligibility}, the SE system development community at large ignores another measure of speech degradation that has been thoroughly studied in relation to hearing loss: \textit{listening effort} (LE). LE is based on the concept that extracting meaning from a degraded signal is cognitively demanding~\cite{Peelle2018}.

LE, like intelligibility, can be measured in an objective manner, but there exist no standardized methods~\cite{Gagne2017}. However, there is a large volume of research showing how it is objectively measurable through a myriad of methods that are psychophysiology- or behavioural-based.

Typical psychophysiological effects are changes in pupil size, skin conductance and EEG signals~\cite{Alhanbali2019}. These methods require specialised equipment for testing, and the right training for operation of this equipment. This creates an obvious barrier for widespread use during SE system development.

Behavioural-based methods, on the other hand, usually only require standard PC equipment. However, they generally rely on a dual-task setup that demands three testing rounds for each processing condition. First, a primary task (often a type of speech-in-noise test) is administered alone, then a second task is administered alone (e.g., tracking a moving target displayed on a computer screen with a mouse, or categorizing digits as even or odd). Finally, the subject is asked to perform both tasks concurrently. LE is then obtained as the difference in performance (often measured as an increased need for time) at the secondary task. Hence, dual-task experiments quickly become extremely resource intensive if there is a need to test different processing conditions.

%The different methods for measuring LE additionally correlate badly~\cite{Alhanbali2019}. When putting all of this together, it is of no surprise that adaptation in the field of SE is lacking. However, LE does have the potential to offer a measure that can really tell whether the SE system has objectively improved the signal (or not) at signal-to-noise ratios (SNRs) where the system will be used. Another advantage of testing at these higher SNRs is that this is where the systems should have an easier job recovering the signal. Moreover, for the SE system developer, the absolute performance that would require correlated metrics is not nearly as important as the ability to compare values obtained with the same method, but for different processing conditions.

The different methods for measuring LE additionally correlate badly~\cite{Alhanbali2019}. When putting all of this together, it is of no surprise that adaptation in the field of SE is lacking. However, LE does have the potential to offer a measure that can really tell whether the SE system has objectively improved the signal (or not). Another advantage of testing LE is that the increase in effort starts well before intelligibility is affected, i.e., at signal-to-noise ratios (SNRs) where the systems should have an easier job recovering the signal~\cite{WinnListeningEffortNot2021}. This is important, since speech intelligibility tests have to use unrealistically poor signals for testing, making the conclusions from such tests less generalisable.

While the major body of work on behavioural measures of LE focuses on dual-task methods, a few studies show that the speech-in-noise test reaction times by themselves already increase when SNRs worsen (and the required LE increases)~\cite{Houben2013, Reinten2020}. Therefore, in this paper, we expand upon this work by proposing a simple method for measuring differences in LE required for speech that has been processed by different SE systems. Unique to our method is that \emph{we filter away incorrect intelligibility test responses}, because those answers may be due to quick guesses adding noise to the measurement. Additionally, we did not inform subjects that they were being tracked for time, and only asked them to focus on understanding the speech. As such, we obtain a relative measure of LE and speech intelligibility with a single test that comes at no extra effort/stress to the tested subjects; a feat that is highly important when every single subject is to repeat the test for multiple processing conditions.

We evaluate the method from the SE system development perspective, and analyse the results from two independent evaluation sessions conducted in Norway and Denmark, respectively. The total number of test participants was 76, and a total of 9 processing conditions were compared.

%Review of listening effort lit:
%\begin{itemize}
%    \item \href{https://journals.lww.com/ear-hearing/FullText/2018/03000/Listening_Effort__How_the_Cognitive_Consequences.4.aspx}{some general comments on cognitive effect of listening effort}
%    \item \href{https://pubmed.ncbi.nlm.nih.gov/24673660/}{review of general listening effort metrics}
%    \item \href{https://journals.lww.com/ear-hearing/fulltext/2019/09000/Measures_of_Listening_Effort_Are_Multidimensional.4.aspx}{measures correlate badly} 
%    \item \href{https://journals.sagepub.com/doi/full/10.1177/2331216516687287}{review of dual task listening effort}
%    \item \href{https://journals.sagepub.com/doi/full/10.1177/23312165211027688}{Listening Effort Is Not the Same as Speech Intelligibility Score}
%    \item \href{https://www.tandfonline.com/doi/abs/10.3109/14992027.2013.832415 }{response time to measure listening effect}
%    \item \href{https://wires.onlinelibrary.wiley.com/doi/full/10.1002/wcs.1514}{cognition or affect?}
%\end{itemize}

\section{Experiments}
\label{sec:experiments}

%The method is based on an adaptation of a commonly used speech recognition test setup that is often referred to as a `matrix test'. During this test, subjects are presented with Hagerman sentences~\cite{Hagerman1995}, which correspond to sequences of 5 mono- and bi-syllabic words according to the syntactical structure \texttt{name} + \texttt{verb} + \texttt{numeral} + \texttt{adjective} + \texttt{object}. Given any of the above 5-word classes, a particular word is drawn from a set of 10 different words in order to create such 5-word sentences. The test interface presents all candidate words for each of the 5-word classes. By a mouse click, a test subject initiates stimulus playback. Then, the test subject picks, by mouse clicks from the candidate word matrix, the words that she/he hears. This procedure is repeated until the test finalizes. The end result is an SRT, which is the SNR where the test subject understands 50$\%$ of the words presented.

The method refines a standard `matrix test’ for speech recognition, utilizing Hagerman sentences~\cite{Hagerman1995}. These sentences are structured as 5-word sequences: \texttt{name}, \texttt{verb}, \texttt{numeral}, \texttt{adjective}, and \texttt{object}, each drawn from a set of 10 words. Subjects interact via mouse to initiate stimulus playback and select perceived words from a matrix. The test concludes with the determination of the word recognition rate (i.e., speech intelligibility).

%an SRT, the SNR at which 50$\%$ of words are understood.

There are several important advantages to matrix tests. First of all, there are 10 possible options for each word class, giving $10^5$ possible unique sentences, all of which are grammatically correct, despite the limited amount of speech material required. This means that matrix tests can be repeated for a same subject without running the risk of the subject being able to answer directly from memory. This makes matrix tests suitable for testing different processing conditions, as is often required when evaluating SE systems. Secondly, subjects quickly become familiar with the speech material, which speeds up the establishment of the learning effect~\cite{Hagerman1995}. Once the subject is familiar with the speech material, the subject's answers only represent the effects of the processing condition to be evaluated. Last but not least, 5-word Hagerman sentences provide realistic listening conditions that are more cognitively challenging than tests based on single words or shorter segments. This means that the LE required should already increase even at low levels of degradation.

%To establish the learning effect, all subjects in both studies were asked to complete a trial round to familiarize themselves with the setup and speech material. After the trial round, subjects were asked to repeat the test for different processing conditions. These processing conditions were presented in random order, to prevent fatigue from repeated rounds to affect the results averaged over subjects. Participants were furthermore encouraged to take breaks between different rounds.

For this study, we extended existing implementations of speech-in-noise tests with timers that track how much time passed between the start of playback of a sentence/stimulus and the subject's click on each of the words. Crucially, subjects were \textit{not} informed about the fact that their reaction times would be tracked. This means that subjects solely focused on the primary task: understanding of speech, and did not experience any additional stress due to a time pressure that is not present in natural listening situations. This is especially important when asking subjects to complete several rounds for different processing conditions. Additionally, it has been stated that LE and speech intelligibility are not the same~\cite{WinnListeningEffortNot2021}. If reaction time can be used to measure LE, the suggested method can give a measure of both speech intelligibility and LE in an easy way. 

%For the analysis, we focused on the reaction time of the first click. This did not necessarily have to be the first word of the sentence, as some participants were observed to adapt to different strategies (like answering in opposite order), especially at lower SNRs. Then we discarded all answers where this first click was on an incorrect word. This was done in order to reduce the `noise' from the quick guesses that many subjects reverted to when the signals became too degraded and incomprehensible.

Two speech-in-noise tests were conducted independently and with different setups. The following two subsections describe how these tests differed.

\subsection{Norwegian Intelligibility Test Description}
\label{ssec:norwegian_setup}

For the Norwegian test, 50 office workers were recruited, with ages ranging from 26 to 72 years old. The recruitment process was purposely inclusive to ensure a wide range of speech recognition thresholds (SRTs). Therefore, the subjects included native Norwegian speakers with or without self-reported/suspected hearing loss, and non-native speakers with self-reported normal hearing, but varying levels of experience with the language. The intelligibility test was repeated for six different processing conditions (models): \textbf{Baseline 1} (noisy), \textbf{Baseline 2} (beamforming with estimated direction), \textbf{Baseline 3} (beamforming with oracle direction), \textbf{SE system 1} (DCCRN), \textbf{SE system 2} (beamforming with estimated direction + DCCRN), and \textbf{SE system 3} (beamforming with oracle direction + DCCRN). Further details of the Norwegian speech intelligibility test are published in~\cite{Gelderblom2024}.
%\begin{itemize}[nosep]
%	\item \textbf{Baseline 1, Noisy}: Single-channel noisy and reverberant speech.
%	\item \textbf{Baseline 2, MPDR (estimated TDOAs)}: Multi-channel noisy and reverberant speech that has been dereverbed with WPE~\cite{Nakatani2010} and beamformed with the minimum power distortionless response (MPDR) beamformer~\cite{VanTrees2002}, where TDOAs were estimated using GCC-PHAT~\cite{Brandstein1997} on the noisy reverberant input.
%	\item \textbf{Baseline 3, MPDR (oracle TDOAs)}: Equal to Baseline 2, but using oracle TDOAs.
%	\item \textbf{SE system 1, DCCRN}: Single-channel noisy and reverberant speech passed through a WPE block and a single-channel deep complex convolutional recurrent network (DCCRN) SE model~\cite{Hu2020, Gelderblom2021, Gelderblom2024}.
%	\item \textbf{SE system 2, MPDR (estimated TDOAs) + DCCRN}: Multi-channel noisy and reverberant speech that has been passed through the combination of Baseline 2 and SE system 1~\cite{Gelderblom2021,Gelderblom2024}. Here, TDOAs were estimated from the dereverberated output of the WPE blocks using GCC-PHAT.
%	\item \textbf{SE system 3, MPDR (oracle TDOAs) + DCCRN}: Equal to SE system 3, but using oracle TDOAs~\cite{Gelderblom2021,Gelderblom2024}.
%\end{itemize}

To find SRTs for the participants, we relied on a Norwegian implementation of a Hagerman test~\cite{Oygarden2009}. The speech material for this test is obtained from a single male speaker. The specific implementation of the intelligibility test used relies on the adaptive psychometric function estimation procedure called the $\Psi$-method~\cite{Prins2009}. This procedure estimates both the threshold and the slope of the psychometric function. 
%Even if the test had a theoretical guess rate of 1/10, the test was not made forced-choice, hence the guess rate was set to 0.01 based on previous experiences from similar tests. Lapse of attention was also set to 0.01. 
The $\Psi$-method also suggests the next stimulus level such that maximum information is added to the system (minimizing the entropy). As such, the subject's answers are not obtained for a fixed set of SNRs, but across a range of SNRs around the subject's personal SRT for a particular processing condition.

During the analysis, the participants were grouped into three categories dependent on participants' SRT results from the unprocessed noisy signal. The grouping was done in such a way that all groups had roughly the same number of participants. A \emph{Low} group ($n=16$) was defined from those with SRT $<-15$ dB, a \emph{Medium} group ($n=17$) for SRT between $-13$ dB and $-15$ dB, and a \emph{High} group ($n=16$) with SRT~$>-13$ dB. 

%The implementation included high resolution timers on all clicks of the user in the answer matrix.

\subsection{Danish Intelligibility Test Description}
\label{ssec:danish_setup}

In total, 26 native Danish speakers (19 males and 7 females, with ages ranging from 18 to 30 years old) took part in the Danish speech intelligibility test. No participant self-reported hearing loss, whereas one subject stated a \emph{slight} bilateral tinnitus. The stimuli presented to test participants can be categorized into three broad processing conditions, namely: \textbf{Unprocessed} (noisy), \textbf{Unmatched} (end-to-end fully-convolutional neural networks trained on a variety of noisy conditions and speakers), and \textbf{Matched} (same as Unmatched, but trained on speech data covering the same noisy condition, language and speaker as those seen at test time).
%\begin{itemize}[nosep]
%	\item \textbf{Unprocessed}: Single-channel noisy speech considering 2 noise types $\times$ 2 SNRs = 4 noisy conditions. The 2 noise types are speech-shaped noise (SSN) and caf\'e, while the 2 SNRs are -10 dB and -5 dB.
%	\item \textbf{Unmatched}: End-to-end (E2E) fully-convolutional neural networks
%	%\footnote{Networks were trained, \emph{using English speech only}, by different perceptually- and non-perceptually-motivated loss functions.} 
%	\cite{Lopez23} were used to enhance single-channel noisy speech as in Unprocessed. These networks correspond to general-purpose enhancement models, which means that they were exposed to a variety of noisy conditions and speakers during the training phase possibly different from those seen at test time.
%	\item \textbf{Matched}: Unlike Unmatched, the Matched processing condition involved the use of E2E fully-convolutional models that were trained/fine-tuned using speech data covering the same noisy condition, language (i.e., Danish) and speaker as those seen at test time.
%\end{itemize}

Note that the speech intelligibility test stimuli were generated from the Dantale II dataset \cite{DantaleII,Wagener03} comprising a single female native Danish speaker. At test time, stimuli playback start time instants as well as time instants of test subjects' button clicks were recorded with a 1-second time resolution.

For further details about the Danish speech intelligibility test the reader is referred to \cite{Lopez23}.

\subsection{Statistical Analysis}
\label{ssec:stat_analysis}

Reaction times typically exhibit non-Gaussian distributions with skewness towards longer durations~\cite{Whelan2008}. Analysing such data challenges traditional Gaussian methods like analysis of covariance (ANCOVA), often diminishing test sensitivity. To reduce the impact of the long tail, elimination of long reaction times from the dataset is a suggested method that can improve the detection of changes in the central tendency~\cite{Ratcliff1993}. The ANCOVA analysis performed on the Norwegian data in this study therefore removed reaction times more than two standard deviations away from the mean value.

Linear regression lines were fitted to the Norwegian data, although it is assumed that the actual shape of reaction times should follow an inverted sigmoid, or even more complicated curves. The argument is that the reaction time will reach a minimum when the SNR is good enough (the person's lowest reaction time), and a maximum at low SNRs (right before the person starts to give up). The reason why we used a linear regression was that the $\Psi$-method mostly tests at challenging SNRs, where the slope of the psychometric function is largest and the persons can recognize some words.

On the Danish data, a Kruskal-Wallis $H$ test was used to compare the conditions. It is a non-parametric variant of the classical parametric one-way analysis of variance (ANOVA). %\cite{ANOVA}
The use of a Kruskal-Wallis $H$ test is motivated by a twofold reason: \emph{1)} we cannot assume Gaussianity for the different reaction time sample populations, but \emph{2)} we can reasonably assume that they follow a similar distribution.

\section{Results}
\label{sec:results}

%We consider all those stimuli for which a subject's first click was correct. Ok, you mentioned this above, so perhaps we can skip it from here or just recall it (?).

\subsection{Norwegian Results}
\label{ssec:norwegian_results}
%Figure~\ref{fig:ANCOVA} shows the regression lines for the reaction times as function of SNR, for each processing condition (Model) and Group.
Linear regression lines were fitted to the data and ANCOVA was performed using SNR, Model (processing condition) and Group (Low, Medium and High) as dependent variables. Table~\ref{tab:ANCOVA} displays the statistical analysis. None of the processing conditions shows any significant effect, but both SNR and Group show highly significant effects.

%\begin{figure*}
%    \centering
%    \includegraphics{Ancova_result.eps}
%    \caption{Regression lines for each model and group.}
%    \label{fig:ANCOVA}
%\end{figure*}

%\begin{table}[t]
%	\centering
%	\begin{tabular}{lccc}
	%			\toprule[1pt]\midrule[0.3pt]
	%			\textbf{Effect} & \textbf{dF} & \textbf{F} & \textbf{p} \\
	%			\midrule
	%			SNR & 1 & 124.212 & $<$0.001 \\
	%           Model & 5 & 1.754 & 0.119 \\
	%          Group & 2 & 92.851 & $<$0.001 \\
	%		\midrule[0.3pt]\bottomrule[1pt]
	%	\end{tabular}
%	\caption{ANCOVA results for the Norwegian dataset.}
%	\label{tab:ANCOVA}
%\end{table}

\begin{table}[t]
	\centering
	\caption{ANCOVA results for the Norwegian dataset.}
	\begin{tabular}{lccc}
		\toprule
		\textbf{Effect} & \textbf{dF} & \textbf{F} & \textbf{p} \\
		\midrule
		SNR & 1 & 124.212 & $<$0.001 \\
		Model & 5 & 1.754 & 0.119 \\
		Group & 2 & 92.851 & $<$0.001 
	\end{tabular}
	\label{tab:ANCOVA}
\end{table}

\begin{figure}
	\centering
	%\includegraphics[width=\linewidth]{RegressionLines.eps}
	%\resizebox{\linewidth}{!}{\input{regressionline}}
	\input{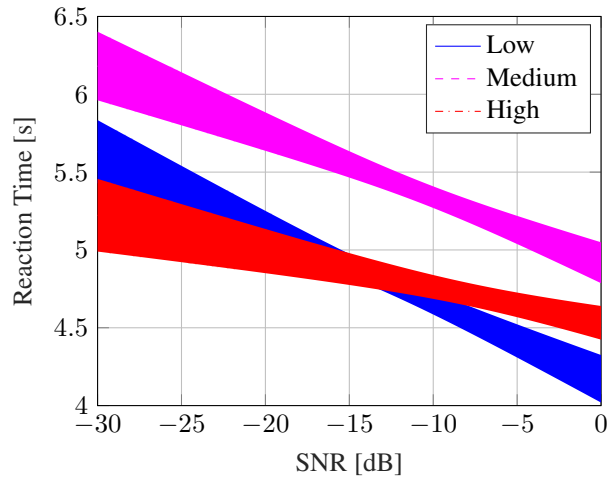}
	\caption{Regression lines with 95\% confidence intervals (shaded areas) for the three groups in the Norwegian dataset.}
	\label{fig:regression}
\end{figure}

To further demonstrate the correlation between reaction time and SNR, Figure~\ref{fig:regression} depicts the regression lines for the different groups. The Low and Medium groups have a similar slope (approx. $-0.45$ s/10 dB), but the Medium group is shifted to the right (statistically significant). The High group has a different slope (statistically significant) from the two other groups, with approx. $-0.23$ s/10 dB.

\subsection{Danish Results}
\label{ssec:danish_results}

%\begin{table}[t]
%	\centering
%	\begin{tabular}{lcc}
	%			\toprule[1pt]\midrule[0.3pt]
	%			\multicolumn{1}{l}{\textbf{Processing condition}} & \multicolumn{2}{c}{\textbf{SNR (dB)}} \\ \cline{2-3}
	%			\multicolumn{1}{l}{} & -10 & -5 \\
	%			\midrule
	%			Unprocessed & 6.02 $\pm$ 0.18 & 5.66 $\pm$ 0.12 \\
	%            Unmatched & 6.18 $\pm$ 0.16 & 5.90 $\pm$ 0.12 \\
	%            Matched & 6.13 $\pm$ 0.16 & 5.62 $\pm$ 0.11 \\
	%			\midrule[0.3pt]\bottomrule[1pt]
	%	\end{tabular}
%	\caption{Mean reaction times, in seconds, with 95\% confidence intervals obtained for the Danish intelligibility test.}
%	\label{tab:times_danish}
%\end{table}

\begin{table}[t]
	\centering
	\caption{Mean reaction times, in seconds, with 95\% confidence intervals obtained for the Danish intelligibility test.}
	\label{tab:times_danish}
	\begin{tabular}{lcc}
		\toprule
		\multicolumn{1}{l}{\textbf{Processing condition}} & \multicolumn{2}{c}{\textbf{SNR (dB)}} \\ \cline{2-3}
		\multicolumn{1}{l}{} & $-10$ & $-5$ \\
		\midrule
		Unprocessed & 6.02 $\pm$ 0.18 & 5.66 $\pm$ 0.12 \\
		Unmatched & 6.18 $\pm$ 0.16 & 5.90 $\pm$ 0.12 \\
		Matched & 6.13 $\pm$ 0.16 & 5.62 $\pm$ 0.11 
	\end{tabular}
\end{table}

%\begin{table}[t]
%	\centering
%	\begin{tabular}{lcc}
	%			\toprule[1pt]\midrule[0.3pt]
	%			\multicolumn{1}{l}{\emph{$p$-values}} & \multicolumn{2}{c}{\textbf{SNR (dB)}} \\ \cline{2-3}
	%			\multicolumn{1}{l}{\textbf{Comparison}} & -10 & -5 \\
	%			\midrule
	%			Unprocessed -- Unmatched & 0.905 & 0.556 \\
	%            Unprocessed -- Matched & 1.000 & 0.765 \\
	%            Unmatched -- Matched & 0.900 & 0.009 \\
	%			\midrule[0.3pt]\bottomrule[1pt]
	%	\end{tabular}
%	\caption{$p$-values from the comparison between pairs of processing conditions obtained for the Danish test.}
%	\label{tab:pval_danish_methods}
%\end{table}

\begin{table}[t]
	\centering
	\caption{$p$-values from the comparison between pairs of processing conditions obtained for the Danish test.}
	\label{tab:pval_danish_methods}
	\begin{tabular}{lcc}
		\toprule
		\multicolumn{1}{l}{\emph{$p$-values}} & \multicolumn{2}{c}{\textbf{SNR (dB)}} \\ \cline{2-3}
		\multicolumn{1}{l}{\textbf{Comparison}} & $-10$ & $-5$ \\
		\midrule
		Unprocessed -- Unmatched & 0.905 & 0.556 \\
		Unprocessed -- Matched & 1.000 & 0.765 \\
		Unmatched -- Matched & 0.900 & 0.009
	\end{tabular}
\end{table}

%\begin{table}[t]
%	\centering
%	\begin{tabular}{lc}
	%			\toprule[1pt]\midrule[0.3pt]
	%			\multicolumn{1}{l}{\textbf{Processing condition}} & \emph{$p$-value} \\
	%			\midrule
	%			Unprocessed & 0.035 \\
	%           Unmatched & 0.019 \\
	%            Matched & $<$0.001 \\
	%			\midrule[0.3pt]\bottomrule[1pt]
	%	\end{tabular}
%	\caption{$p$-values from the comparison between -10 dB and -5 dB SNRs obtained for the Danish test.}
%	\label{tab:pval_danish_snrs}
%\end{table}

\begin{table}[t]
	\centering
	\caption{$p$-values from the comparison between $-10$ dB and $-5$ dB SNRs obtained for the Danish test.}
	\label{tab:pval_danish_snrs}
	\begin{tabular}{lc}
		\toprule
		\multicolumn{1}{l}{\textbf{Processing condition}} & \emph{$p$-value} \\
		\midrule
		Unprocessed & 0.035 \\
		Unmatched & 0.019 \\
		Matched & $<$0.001 
	\end{tabular}
\end{table}

Grouped across speech-shaped and caf\'e noises, Table \ref{tab:times_danish} depicts test subjects' mean reaction times along with 95\% confidence intervals as a function of the SNR and processing condition. Given these mean reaction times, Table \ref{tab:pval_danish_methods} reports $p$-values from the comparison between processing condition pairs. As can be seen, only at $-5$ dB we can observe a statistically significant difference between Unmatched and Matched. Similarly, Table \ref{tab:pval_danish_snrs} shows $p$-values from the comparison between $-10$ dB and $-5$ dB SNRs given the different processing conditions. As we can see from this table, given \emph{any} processing condition, $-5$ dB mean reaction times are statistically significantly shorter with respect to those at $-10$ dB.

\section{Discussion}
\label{sec:discussion}

The Norwegian study tested multiple SNRs, making it possible to search for a correlation between SNR and reaction time. None of the processing conditions tested gave any significant changes in reaction time, but the groups showed a statistically significant difference. 

Two groups (Low and Medium) showed similar slopes (approx. $-0.45$ s/10 dB), but the curve for the Medium group was shifted to the right (longer reaction times). This can be interpreted as if the Medium group had an increased LE at the same SNR as the Low group (those with best hearing). The High group (those with worst results on the unprocessed signal) had a different slope. A possible interpretation of the regression line is that they struggle as much as the Medium group at high SNRs, but that they quickly give up. It should be noticed that the High group was very inhomogeneous, and many were struggling to understand all of the words, even at the best SNRs.
%The standard deviation of the SRTs of the High group was much higher than the two other ($\text{SD}_\text{Low}=0.8~\text{dB}$, $\text{SD}_\text{Medium}=0.5~\text{dB}$, and $\text{SD}_\text{High}=2.9~\text{dB}$). 
Therefore, the results from the High group should be treated with care.

A limitation in our analysis is that we have only done a linear regression on the data. It is assumed that the reaction time (as function of SNR) follows an inverted sigmoid shape, or even more complicated curves. Future studies should investigate this further. Fitting a linear regression to data that actually follow an inverted sigmoid will give a less steep slope. Since the results still showed a statistically significant trend, a more complicated fitting would have given even stronger evidence. A future study should investigate the shape of the distribution, testing over a larger range of SNRs.

The Danish test only evaluated LE at two SNRs, but found statistically significant changes between the two. Converting the changes into a slope gave between $-0.56$ s/10 dB and $-1.02$ s/10 dB, slightly higher than the Norwegian study. This could be because the two SNRs tested were at levels where the slope actually was higher. Besides, the statistically significant difference between Unmatched and Matched at $-5$~dB SNR might be explained by the higher LE required to understand, \emph{due to increased speech distortion}, the  stimuli processed by models trained with data from acoustic conditions distinct from those found at test time.

%The highest slope was also found on the processed signal (Matched). This can be interpreted as if the processing made the signal more difficult to understand and required a higher LE at -10 dB SNR.
%This comes as no surprise, since test subjects were having a harder time figuring out, \emph{due to increased speech distortion}, the spoken words in the stimuli processed by enhancement models trained with data from acoustic conditions distinct from those found at test time.

Another limitation in the Danish study was the time resolution used for reaction time estimation. Since the reaction time never was intended to be measured, the time could only be found with a 1-second resolution. Nonetheless, the changes in reaction time still were statistically significant, but the resolution could be a reason why we see differences between the two studies.

\section{Conclusion}
\label{sec:conclusion}

Reliable ways of evaluating SE systems are important. LE is one way of doing this, but no standardized method exists, and a common issue is the complicated methodology that creates barriers for performing such evaluation. In our studies, a simple intelligibility test (5-word Hagerman sentences) was performed, where the test subjects clicked on the words they heard in a graphical user interface, with 10 words in each of the 5 categories. The reaction time to the first (correct) answer was collected and defined as the reaction time of the subject.

Both the Norwegian and the Danish test showed clear statistically significant changes in reaction time for different SNRs. This is a clear indication that the test subjects needed more or less time to understand what was said, and that the LE changed. An important point to notice is that the increase in reaction time begins before intelligibility is affected. This means that reaction time can be used as an evaluation metric at SNRs that are more realistic in, for instance, video conferencing, podcasts and other audio productions. Using the suggested method also makes it possible to measure both intelligibility and LE in one test. The simplicity of the method (which relies neither on specialized equipment nor on dual-task designs) makes this a candidate for further evaluation.

\bibliographystyle{IEEEbib}
\bibliography{refs}

\begin{thebibliography}{10}

\bibitem{Taal2011}
C.~H. Taal, R.~C. Hendriks, R.~Heusdens, and J.~Jensen,
\newblock ``An {{Algorithm}} for {{Intelligibility Prediction}} of
  {{Time-Frequency Weighted Noisy Speech}},''
\newblock {\em IEEE/ACM Trans. Audio Speech Lang. Process.}, vol. 19, no. 7,
  pp. 2125--2136, Sept. 2011.

\bibitem{ITU2001}
ITU,
\newblock ``Recommendation {{P}}.862: {{Perceptual}} evaluation of speech
  quality ({{PESQ}}): {{An}} objective method for end-to-end speech quality
  assessment of narrow-band telephone networks and speech codecs,'' 2001.

\bibitem{Lopez23}
I.~{L{\'o}pez-Espejo}, A.~Edraki, W.-Y. Chan, Z.-H. Tan, and J.~Jensen,
\newblock ``On the deficiency of intelligibility metrics as proxies for
  subjective intelligibility,''
\newblock {\em Speech Com.}, vol. 150, pp. 9--22, 2023.

\bibitem{Gelderblom2017}
F.~B. Gelderblom, T.~V. Tronstad, and E.~M. Viggen,
\newblock ``Subjective {{Intelligibility}} of {{Deep Neural Network-Based
  Speech Enhancement}},''
\newblock in {\em {{INTERSPEECH}}}, Stockholm, Sweden, Aug. 2017, pp.
  1968--1972, ISCA.

\bibitem{Gelderblom2019}
F.~B. Gelderblom, T.~V. Tronstad, and E.~M. Viggen,
\newblock ``Subjective {{Evaluation}} of a {{Noise-Reduced Training Target}}
  for {{Deep Neural Network-Based Speech Enhancement}},''
\newblock {\em IEEE/ACM Trans. Audio Speech Lang. Process.}, vol. 27, no. 3,
  pp. 583--594, Mar. 2019.

\bibitem{Gelderblom2024}
F.~B. Gelderblom, T.~V. Tronstad, T.~Svendsen, and T.~A. Myrvoll,
\newblock ``On the {{Predictive Power}} of {{Objective Intelligibility
  Metrics}} for the {{Subjective Performance}} of {{Deep Complex Convolutional
  Recurrent Speech Enhancement Networks}},''
\newblock {\em IEEE/ACM Trans. Audio Speech Lang. Process.}, vol. 32, pp.
  215--226, 2024.

\bibitem{Zhao2018}
Y.~Zhao, D.L. Wang, E.~M. Johnson, and E.~W. Healy,
\newblock ``A deep learning based segregation algorithm to increase speech
  intelligibility for hearing-impaired listeners in reverberant-noisy
  conditions,''
\newblock {\em J. Acoust. Soc. Am.}, vol. 144, no. 3, pp. 1627--1637, Sept.
  2018.

\bibitem{Healy2015}
E.~W. Healy, S.~E. Yoho, J.~Chen, Y.~Wang, and D.L. Wang,
\newblock ``An algorithm to increase speech intelligibility for
  hearing-impaired listeners in novel segments of the same noise type,''
\newblock {\em J. Acoust. Soc. Am.}, vol. 138, no. 3, pp. 1660--1669, 2015.

\bibitem{Lim1979}
J.~S. Lim and A.~V. Oppenheim,
\newblock ``Enhancement and bandwidth compression of noisy speech,''
\newblock {\em Proceedings of the IEEE}, vol. 67, no. 12, pp. 1586--1604, 1979.

\bibitem{Peelle2018}
J.~E. Peelle,
\newblock ``Listening {{Effort}}: {{How}} the {{Cognitive Consequences}} of
  {{Acoustic Challenge Are Reflected}} in {{Brain}} and {{Behavior}},''
\newblock {\em Ear \& Hearing}, vol. 39, no. 2, pp. 204--214, Mar. 2018.

\bibitem{Gagne2017}
J.-P. Gagn{\'e}, J.~Besser, and U.~Lemke,
\newblock ``Behavioral {{Assessment}} of {{Listening Effort Using}} a
  {{Dual-Task Paradigm}}: {{A Review}},''
\newblock {\em Trends in Hearing}, vol. 21, pp. 233121651668728, Dec. 2017.

\bibitem{Alhanbali2019}
S.~Alhanbali, P.~Dawes, R.~E. Millman, and K.~J. Munro,
\newblock ``Measures of {{Listening Effort Are Multidimensional}},''
\newblock {\em Ear \& Hearing}, vol. 40, no. 5, pp. 1084--1097, Sept. 2019.

\bibitem{WinnListeningEffortNot2021}
M.~B. Winn and K.~H. Teece,
\newblock ``Listening {{Effort Is Not}} the {{Same}} as {{Speech
  Intelligibility Score}},''
\newblock {\em Trends in Hearing}, vol. 25, pp. 233121652110276.

\bibitem{Houben2013}
R.~Houben, {M. van Doorn-Bierman}, and {W. A. Dreschler},
\newblock ``Using response time to speech as a measure for listening effort,''
\newblock {\em Int. J. Audiol.}, vol. 52, no. 11, pp. 753--761, 2013.

\bibitem{Reinten2020}
I.~Reinten, I.~{de Ronde-Brons}, M.~{van den Tillaart-Haverkate}, R.~Houben,
  and W.~Dreschler,
\newblock ``Using response times to speech-in-noise to measure the influence of
  noise reduction on listening effort,''
\newblock {\em Proc. ISAAR}, vol. 7, pp. 165--172, 2020.

\bibitem{Hagerman1995}
B.~Hagerman and C.~Kinnefors,
\newblock ``Efficient adaptive methods for measuring speech reception threshold
  in quiet and in noise,''
\newblock {\em Scand. Audiol.}, vol. 24, no. 1, pp. 71--77, 1995.

\bibitem{Oygarden2009}
J.~{\O}ygarden,
\newblock {\em Norwegian Speech Audiometry},
\newblock Ph.D. thesis, Norwegian University of Science and Technology (NTNU),
  Trondheim, Norway, 2009.

\bibitem{Prins2009}
N.~Prins and F.~A.~A. Kingdom,
\newblock ``Palamedes: {{Matlab}} routines for analyzing psychophysical data,''
  \url{http://www.palamedestoolbox.org}, 2009.

\bibitem{DantaleII}
M.~Hansen and C.~Ludvigsen,
\newblock ``Dantale {{II}}: {{Danske Hagerman}} s{\ae}tninger,''
\newblock Tech. {R}ep., 2001.

\bibitem{Wagener03}
K.~Wagener, J.~L. Josvassen, and R.~Ardenkj{\ae}r,
\newblock ``Design, optimization and evaluation of a {{Danish}} sentence test
  in noise: {{Dise{\~n}o}}, optimizaci{\'o}n y evaluaci{\'o}n de la prueba
  {{Danesa}} de frases en ruido,''
\newblock {\em Int. J. Audiol.}, vol. 42, pp. 10--17, 2003.

\bibitem{Whelan2008}
R.~Whelan,
\newblock ``Effective {{Analysis}} of {{Reaction Time Data}},''
\newblock {\em The Psychol. Rec.}, vol. 58, no. 3, pp. 475--482.

\bibitem{Ratcliff1993}
R.~Ratcliff,
\newblock ``Methods for dealing with reaction time outliers,''
\newblock {\em Psychol. Bull.}, vol. 114, no. 3, pp. 510--532.

\end{thebibliography}

\end{document}